# Strategies for Addressing Spreadsheet Compliance Challenges

Microsoft Corporation
1 Microsoft Way. Redmond WA 98052
bweber@microsoft.com

## ABSTRACT

*Most organizations today use spreadsheets in some form or another to support critical business processes. However the financial resources, and developmental rigor dedicated to them are often minor in comparison to other enterprise technology. The increasing focus on achieving regulatory and other forms of compliance over key technology assets has made it clear that organizations must regard spreadsheets as an enterprise resource and account for them when developing an overall compliance strategy. This paper provides the reader with a set of practical strategies for addressing spreadsheet compliance from an organizational perspective. It then presents capabilities offered in the 2007 Microsoft Office System which can be used to help customers address compliance challenges.*

## 1. INTRODUCTION

Because of their ease of use, flexibility, and power, spreadsheets support many critical business functions, and often fill roles where other solutions would be too slow or costly to implement. As a result, they have quietly become a key component in the analysis and reporting processes within most organizations, including the mission critical area of financial reporting.

In the US and around the world, there has been increasing focus on demonstrating regulatory compliance especially within corporate financial processes. This is due in part to new legislation, such as Sarbanes-Oxley, as well as an increased public scrutiny of corporate accounting practices, and has highlighted a need for stricter controls over the analysis supporting financial statements. Because spreadsheets are an enterprise resource that support key business processes, it is important to determine how they fit into an overall strategy for regulatory compliance.

This paper will provide the reader with a set of practical strategies for addressing spreadsheet compliance from both an organizational and technological level. While these strategies focus on financial analysis and reporting scenarios, they are not regulation specific and can be applied to spreadsheet environments across domains and industries.

## 2. REGULATORY COMPLIANCE

### 2.1. Background

Regulatory compliance is now, more than ever, a top of mind issue for organizations around the world. One area in particular that has received much scrutiny in recent years is financial compliance, where new legislation has been written to ensure that organizations' financial analysis and reporting processes are both transparent and accurate. The three most visible examples of this legislation are the Sarbanes-Oxley Act (United States, 2002), Data Protection Act (European Union, 1998) and the Basel Capital Accord (Basel II, 2006) which together, affect most publicly traded companies. Corporate finance, however, is just one of many areas where compliance policies have been defined and enforced. The pharmaceutical and health industries, for example, have been subject to strict regulation for years. Compliance policies may also vary based on the location of the organization, with regional governments often requiring their own set of controls. An organization must take into account all applicable policies and requirements when developing a regulatory compliance framework.

### 2.2. Spreadsheets: An enterprise software resource

Though they may not be thought of in the same manner as database or custom software systems, spreadsheets are a key enterprise asset for most organizations. In the words of auditor PricewaterhouseCoopers, spreadsheets are "an integral part of the information and decision-making framework for companies"[i]. However, evidence has shown that in some organizations there is the general perception that spreadsheets are a tactical tool without strategic importance. As a result, the resources dedicated toward the implementation and control of critical spreadsheets are small in comparison to other information technology assets. These disparities represent the most significant road block to spreadsheet compliance. Before controls can be implemented and enforced, management must acknowledge spreadsheets as a critical enterprise resource then budget and plan accordingly.

### 2.3. A process challenge

One common misconception in organizations is that the solution for spreadsheet compliance is technology. While technology plays a role in any compliance strategy, the most important component is process. Critical spreadsheets, like other enterprise IT resources, require sound development and usage practices that include controlled testing, deployment, maintenance, and use. An effective plan will incorporate these steps into the larger compliance framework for spreadsheets and other enterprise resources. The points listed below are prerequisite to developing such a plan.

#### Executive buy-in

The need for compliance policies must be recognized at all levels within the organization. Cooperation between the management of functional departments is important to defining robust controls that are in line with business objectives. Without executive-level commitment, the collaboration needed to create a broadly enforceable compliance strategy will be difficult if not impossible to achieve.

#### Getting IT and business users on the same page

A compliance strategy must take into account the needs of the business and its users in order to be successful. A plan that fails to do so will come at the expense of business productivity and not be sustainable. This problem can be avoided by engaging users from the start when developing a compliance plan. User input is particularly important when defining spreadsheet controls, as members of the business team often serve as both the user and developer of critical spreadsheet applications.

#### Allocate appropriate resources

Implementing an effective compliance strategy takes time and effort. Human resources must be allocated from many different groups in order to define controls that meet business objectives. These often include the IT, internal audit, and finance departments, but can involve others depending on the needs of the organization. Additionally, it may make sense to use software to facilitate the monitoring and controlling of spreadsheets, which can require financial and development resources as well. Once implemented, the control processes must be monitored and enforced by dedicated people with an understanding of the compliance strategy.

#### Every Situation is Unique

Every organization is unique in how they use technology in their business, and each has its own set of challenges and goals to consider when developing a compliance framework. As a result there is no single prescriptive compliance solution that satisfies the needs of all. An effective strategy will take into account the operational requirements, business objectives, and specific regulations with which the organization must comply. The following section offers some specific steps for identifying and controlling critical financial spreadsheets, but these strategies can be adapted to meet the particular needs of the organization's larger compliance framework.

## 3. COMPLIANCE STRATEGIES FOR SPREADSHEETS

Once an organization's overall framework for regulatory compliance is in place, the following steps can be taken to identify and control critical spreadsheets in a way that can be maintained into the future. The 3 key steps to implementing this process are:

1. Evaluate the current situation
2. Implement the appropriate controls
3. Develop a long term spreadsheet development and maintenance methodology

This section will cover each of these in detail.

### 3.1. Evaluate the current situation

**Inventory relevant spreadsheets**

Before critical spreadsheets can be controlled they must first be identified. An inventory should be performed to count the population of spreadsheets in the organization which may impact compliance. In most organization's this applies to spreadsheets concentrated in specific high risk departments. In the case of corporate finance it would include areas supporting the analysis and reporting of financial accounting data. During the inventory each spreadsheet's purpose and relationship to critical business processes should be noted. This information will be important when later determining the appropriate controls for each spreadsheet. Inventories can be carried manually by inspecting hard drives and shared folders, or in a more automated fashion using software that scans a corporate network to target spreadsheets.

**Identify the business-critical spreadsheets**

Not all spreadsheets in an organization require rigorous compliance controls. An inventory may return thousands, if not hundreds of thousands of spreadsheets, many of which will likely have little or no compliance impact. It would be overwhelming and unproductive for an organization to implement strict controls for each spreadsheet found in their enterprise. As a result, teams must identify and isolate the spreadsheets which support critical business processes, where a lack of controls could lead to material errors. In corporate finance, a material error is typically defined as one that impacts 5% of the general ledger, but this definition should be tailored to satisfy the organization's compliance requirements and departments involved. In most cases only a small percentage of an organization's spreadsheets match these criteria.

**Case Study: Microsoft Corporation**

As an example, Microsoft Corporation's Financial Compliance Group works with management, Internal Audit, and the External Auditors to perform an inventory of important spreadsheets used for financial reporting. A recent inventory yielded a total of 42 "business critical" spreadsheets in use. The following filtering criteria were used to identify these spreadsheets.

<u>Review criteria</u>

Microsoft uses the following criteria to determine which inventoried spreadsheets need to be reviewed for further analysis.

1. A spreadsheet which documents a journal entry greater than a pertinent dollar threshold. This threshold is derived as a percentage of materiality on a quarterly basis to support quarterly reporting.

2. A spreadsheet that serves as a recording ledger for an account with a balance greater than a pertinent dollar threshold. This threshold is derived as a percentage of materiality roughly four times greater than the threshold for supporting a journal entry.
3. A spreadsheet that directly supports a financial statement disclosure.

Control criteria

Control criteria are used to determine whether or not a spreadsheet that passes through the Review filter should be subject to control activities. These criteria are formed by an assessment based on the inherent and historical risk of the information contained within that spreadsheet. A risk assessment was conducted to identify the important risks pertaining to spreadsheets.

Examples of control criteria used by Microsoft include:

1. Complexity of the spreadsheet: a high degree of formula and calculation complexity, connections to multiple worksheets or external data sources, the use of macros and other code, etc.
2. Whether or not the spreadsheet is well documented with an established history of changes made to it.

### 3.2. Implement the appropriate controls

Once business critical spreadsheets have been identified and their risks defined, the next step is to implement the appropriate controls for each one. Often times this process cannot be implemented for the entire organization at once. In this case it's necessary to break the work up by division or business unit, addressing the most important areas first. The section below describes the risks and corresponding control activities Microsoft uses to control its critical spreadsheets.

**Control Activities**

Control activities are the actual steps taken to mitigate risk and meet the control objective. The control activities in use by Microsoft's corporate controllers fall into 4 categories:

- **Preventative** – Controls that prevent undesirable events from occurring
- **Detective** – Controls that detect undesirable events which have already occurred
- **Directive** – Controls that cause or encourage a desirable event to occur
- **Mitigating** – Balancing employee empowerment and cost with alternative controls

The table below specifies the potential compliance risks for business-critical spreadsheets and their corresponding control activities as defined by Microsoft's Financial Compliance Group.

| Potential Spreadsheet Risk | Control Activity | Category |
|---|---|---|
| **Unauthorized access and modification of data or formulas may result in output / reporting error. Loss of the archived (prior period) spreadsheets may damage audit trail.** | Save spreadsheets to a location that enables restricted user access & provides regular back-ups. | Preventative |
| **Unauthorized modification of historical data may damage audit trail.** | Spreadsheets from previous reporting periods should be converted to a "read-only" format and securely archived for later retrieval. | Detective |

| Potential Spreadsheet Risk | Control Activity | Category |
|---|---|---|
| **Spreadsheets may be initially set-up with incorrect formulas. Informal or inadvertent spreadsheet changes subsequent to development may degrade model integrity which results in output & reporting errors.** | 1. Overall model mechanics are tested before spreadsheet is used. Material changes are tested before spreadsheet is used. Spreadsheets are re-tested once a year. | Preventative |
| | 2. Formula cells are locked to prevent inadvertent changes. | Preventative |
| | 3. Mechanics are reviewed every reporting period in sufficient detail to detect inadvertent changes. | Preventative Mitigating |
| **Input data is not entered in its entirety and/or does not agree to the source which may result in output / reporting error.** | Check cells are used to validate data accuracy and the completeness of inputs. | Detective Mitigating |
| **Inability to transfer knowledge regarding how to properly use the spreadsheet degrades current and future ability to generate correct results and accurate reports** | Spreadsheets are divided into three worksheets to separate Input Values, Formulas and Resulting calculations. | Directive |
| | Key elements of spreadsheets are documented: input cells, formula cells, output cells, data sources, calculation methodology summary, and spreadsheet use procedure summary. | Directive |

Defining and implementing control activities, including those described above, is an important step in addressing the compliance risks associated with business critical spreadsheets. The next section describes how an organization can apply these principles to new business critical spreadsheets by implementing a controlled development process.

### 3.3. Develop a long term spreadsheet development and maintenance methodology

In addition to the control processes described above, a long-term strategy for mitigating spreadsheet risk should include the use of a development methodology for critical spreadsheets. Academic research indicates that spreadsheet development shares many of the same characteristics as traditional software development[ii]. Error rates tend to be similar, as do the benefits that can be gained from a sound development lifecycle that includes design, inspection and maintenance. Historically, spreadsheets have not received the level of developmental rigor given to other forms of enterprise software. As a result, spreadsheets driving key aspects of the business often lack important controls and thus introduce a compliance risk for the organization. The solution here is to treat business-critical spreadsheets like enterprise software and adhere to a formal development methodology when creating them.

Below is a recommended development approach to creating spreadsheets[iii]:

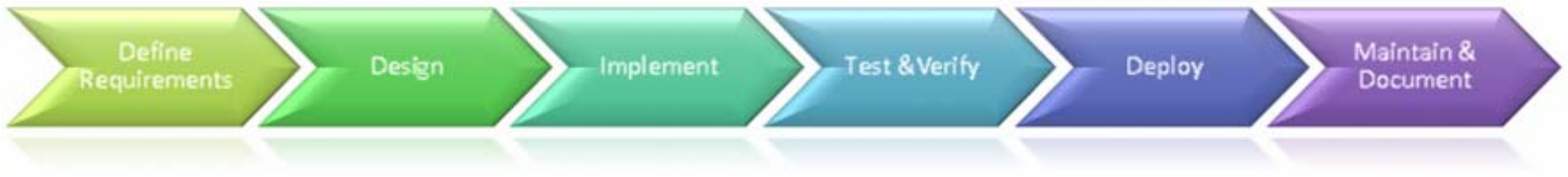

**Define Requirements**

The development of a spreadsheet model should first begin by defining its requirements. This phase should include a detailed description of the spreadsheet's business purpose, including the functions it will perform and its impact on the broader business process. It is wise to scope and define boundaries here as well, as this will help prevent the spreadsheet from growing large and unwieldy during the design and development phases. Additionally, this phase should include the sign-off of spreadsheet users to ensure that the application will satisfy their business needs.

**Design**

The design phase maps out a detailed plan for implementing the business requirements defined in the first phase, and should result in a spreadsheet "blueprint." This blueprint should describe the formulas and functions needed for the core logic as well as the layout of spreadsheet itself. Well-designed spreadsheets include the following characteristics:

- A clear, visual separation of inputs, outputs and calculation cells
    - This can be achieved through layout and placement as well as through formatting
- Locking and protecting cells that should not be modified
- The use of a standard organizational method
    - One common example is the top-down organization where formulas never refer to the cells located below them
- The use of standard naming conventions throughout the spreadsheet
- The use names to reduce errors and increase the readability of formulas
- The use of simple formulas
    - This can be achieved by breaking complex business logic into multiple cells
- Extensive documentation throughout the spreadsheet
    - This is especially effective when comments are embedded throughout the spreadsheet
    - This might also include tables of contents and formatting legends to clarify the structure and layout of the spreadsheets

Again, it is imperative to incorporate spreadsheet user feedback in the Design phase to ensure that the final "blueprint" is flexible enough to be used, but strict enough to respond to organizational controls.

**Implement**

Once the "blueprint" has been created and validated, it is time to create the spreadsheet. If the Requirements and Design phases have been completed with care and a high level of detail, this step should simply assemble the pieces as described in the spreadsheet blueprint.

**Test & Verify**

Like any new piece of custom software, spreadsheets will contain errors. As a result testing and verification of the spreadsheet's calculation accuracy is critical to ensuring confidence in the model itself. There are a number of different ways to 'test' a spreadsheet including targeted audits, test case verification, scenario testing, and code inspection. Of these testing methods code inspection has been shown to be the most complete for catching errors. Research indicates that code inspections tend to find on average over 80% of the errors in spreadsheets[1]. However, it is also the most resource intensive, involving teams of 1-3 reviewers with a firm understanding of the spreadsheet to analyze it closely for logic and input errors. Regardless of the method, test passes should happen regularly throughout the implementation phase by individuals other than those that initially created the spreadsheet. In addition to good testing practices, 3rd party testing tools can be used to help identify and fix spreadsheet errors.

**Deploy**

When deploying, the owner must determine and apply the controls needed for the particular spreadsheet. Examples of spreadsheet controls were given in the previous section of this paper. The controls needed for compliance will vary depending on the complexity and importance of the spreadsheet. Other activities to considering when deploying are:

- A Formal transition to a production environment
    - Source files are backed up
    - Solution is stored in a secure location, with strong file access controls
    - Sign-off from development, test, and business users
- Training and education for users of the solution
    - Creation of a detailed user manual
    - Training courses that educate users and verify proficiency with the solution.

**Maintain & Document**

Maintenance and documentation are critical to the longevity of a spreadsheet, without these its life span will be limited and its overall value reduced. Continued testing and verification of all changes made to the spreadsheet after it is deployed will help ensure that the logic remains correct. Documentation is important as it allows users, developers, and testers to understand the purpose and function of the spreadsheet. This will reduce the amount of future testing needed and minimize user error. Documentation for critical spreadsheets should include the following elements

- A detailed description of the spreadsheet's purpose
- A log of changes made to the spreadsheet that include who made them and how they affected the spreadsheet
- An explanation of input cells using cell comments
    - Description of all data inputs
    - Description of formula cell calculations using cell comments
    - Any standard, defined spreadsheet naming conventions
- A legend that explains the formatting used in the spreadsheet
- A user's manual that explains the proper use of the spreadsheet with example input and output values.
- Contact information for the person who created and is responsible for the spreadsheet.

## 4. HOW THE 2007 MICROSOFT® OFFICE SYSTEM CAN HELP ADDRESS COMPLIANCE CHALLENGES

While technology alone cannot ensure spreadsheet compliance, organizations should take full advantage of the tools and technology available to help fulfill the compliance recommendations outlined above. Risks to spreadsheet compliance can be mitigated by implementing controls on important elements of business-critical spreadsheets, so that only authorized users are able to view content, make changes, and share information.

---

[1] Panko, Raymond R., Ordway Nicholas. *"Sarbanes-Oxley: What about All the Spreadsheets?"* University of Hawaii, 2005

This section presents a set of technologies included in the 2007 release of the Microsoft Office System that can be used in conjunction with a sound compliance strategy to address compliance challenges regarding spreadsheets, including:

- Preventing unauthorized access to spreadsheets
- Managing & monitoring spreadsheet changes
- Retaining & archiving spreadsheets
- Developing robust spreadsheet models

Some of the capabilities that will be described are available in the curent release of the Microsoft Office System.

### 4.1. Preventing unauthorized access to spreadsheets

As the complexity and importance of a spreadsheet increases, so to does the cost of errors and innaproriate disclosures of data. The 2007 Microsoft Office systems offers a number of different options for securing critical spreadsheets on both the client and server from unauthorized access and modification. This section will take a closer look at the following four technologies.

1. Microsoft Office Sharepoint® Server 2007 permissions
2. Sharing Spreadsheets Using Excel Services 2007
3. Information Rights Management
4. Workbook encryption in Excel 2007

#### 4.1.1. Office SharePoint Server 2007 Permissions

Office SharePoint Server 2007 is a scalable enterprise portal, content management, and collaboration server built on Microsoft Windows® SharePoint Services. Organizations can use Office SharePoint Server 2007 to store, protect, share, and track important documents and information through a single Web-based portal. All interactions within Office SharePoint Server 2007 are protected and monitored by a single sign-on system to safeguard against unauthorized access to critical documents.

Office SharePoint Server 2007 uses a security model based on site groups and rights. Site groups are groups of users with related security requirements. Site owners can assign Security rights to each security group. An organization can customize the rights assigned to these site groups or add new site groups as needed. By default, Office SharePoint Server 2007 includes six site groups: Administrator, Web Designer, Contributor, Reader, Guest, and Viewer.

Once groups and permissions have been defined, Office SharePoint Server 2007 safeguards the sites and documents stored within the portal using this permission structure.

#### 4.1.2. Sharing Spreadsheets Using Excel Services

Excel Services is a new server-based technology that supports loading, calculating, and rendering Microsoft Office Excel spreadsheets in a Web browser. Excel Services comprises two primary interfaces: Microsoft Office Excel Web Access allows customers to view spreadsheets in a Web browser and the Excel application programming interface (API) allows developers to share Excel features among applications. With the Microsoft Office system, customers can publish spreadsheets and view them with any modern browser, without the need to install software on the local computer. This allows organizations to share spreadsheets without exposing sensitive business logic. Finally, because Excel Services is part of Office SharePoint Server 2007, it takes full advantage of document management and workflow capabilities to help maintain control over critical spreadsheets.

*Controlling What Users Can See*

Publishing a spreadsheet to Office SharePoint Server 2007 saves the entire spreadsheet to the server to allow for data refreshes and recalculation. However, the parts of the spreadsheet accessible to viewers and available for download through the Web browser are controlled by the author of the spreadsheet.

Microsoft Office Excel 2007 spreadsheet software provides three options for controlling the viewable area of the spreadsheet on the server:

- The entire workbook (default). Users can view the entire workbook and download it to the desktop.
- A subset of sheets. The workbook author permits users to view and download a subset of sheets. This does not affect how the spreadsheet appears when opened in Office Excel 2007, only how it appears when viewed on the server. This mode is useful when workbooks contain numerous “behind the scenes” worksheets that hold intermediate calculations, source data, etc., but only a few sheets that users should see.
- A set of named items, such as Named Ranges, charts, tables, and PivotTable® and PivotChart® dynamic views. In this mode, users can only view and download specific items selected by the workbook author. Users access these items through a drop-down menu in their Web browser.

*The View Item Right*
Office SharePoint Server 2007 adds a new feature for spreadsheets (and other documents) stored in SharePoint document libraries. With this View Item Right, spreadsheet administrators can restrict user access to viewing and executing on the server. Users cannot download a copy of the spreadsheet or access any areas that were not published to be viewable on the server. This feature can hide and make inaccessible proprietary information contained in the workbook, such as specific formulas, the proprietary model, the external data connections, and hidden elements. The View Item Right affects the way Excel Web Access and the Excel API allow access to a workbook.

### 4.1.3. Information Rights Management

Organizations can use Information Rights Management (IRM) to protect and maintain greater control over their digital information, including confidential and sensitive spreadsheets. IRM relies on Microsoft Windows Rights Management Service (RMS) technology in Microsoft Windows Server 2003. RMS must be installed in an Active Directory domain in which the domain controllers are running Windows Server 2000 with Service Pack 3 (SP3) or later.

By taking advantage of IRM, organizations and individual users can set policies that set greater control over who can open, copy, print, or forward information created in Excel 2007.

**IRM in Office Excel 2007**
With IRM, users can set different levels of file protection—balancing the needs to efficiently share information and help protect privacy.

- Set file permissions at different levels and change the level for specific users and groups of users.
- Assign permissions according to roles and responsibilities. For example, set different permissions for a viewer, a reviewer, or a file editor.
- Restrict file printing to reduce the number of printed copies of a sensitive spreadsheet that might be produced.
- Set expiration dates to provide a time limit after which a spreadsheet file can no longer be opened or used by others.
- Help prevent forwarded files from being opened by an unauthorized recipient. Unintended recipients cannot open files protected with IRM. Instead, a message informs them that they do not have access rights. Optionally, the file owner can include an e-mail address.

**IRM and Office SharePoint Server 2007**

Sharepoint document libraries are also highly integrated with Information Rights Management. Using IRM, Sharepoint can apply policy to protect spreadsheets automatically as they are downloaded to a user's laptop. Offline use is unhindered, but as needed rights such as forwarding the spreadsheet, printing, or editing can be disallowed on a user-by-user basis. Finally, Sharepoint Server 2007 can leverage IRM to expire content after a specified time. This helps reduce the problem of having out-of-date versions of a spreadsheet floating around in email, causing confusion.

### 4.1.4. Workbook Encryption in Excel 2007

Customers without Sharepoint Server 2007 deployed can use the Password Protect workbook functionality in Excel 2007 to get a basic level of file security. Password Protect workbook allows users to specify a password required in order to open the workbook. The password is encrypted using a symmetric encryption routine known as 40-bit RC4.

## 4.2. Managing & Monitoring Spreadsheet Changes using SharePoint Server 2007

Critical Spreadsheets are living applications that inevitably change over time. A sound compliance strategy will include some level of on-going change management and monitoring for critical spreadsheets. In this section we will take a closer look three new capabilities in SharePoint Server 2007 that allow customers to better manage the important spreadsheets and documents in their organization without sacrificing productivity.

**Versioning**

Office SharePoint Services has a robust check-in/check-out and versioning mechanism that allows users to check in changes under a new major (1.0 to 2.0) or minor (1.8 to 1.9) version. Sharepoint will keep as many back versions as is needed with full version history showing who and when each version was created.

**Auditing**

Office System SharePoint Server 2007 allows administrators to audit key events within document libraries. While there is no built-in capability to audit the changes within spreadsheets themselves, events such as Open, Create, Modify, and Delete, of spreadsheets are all logged to a centralized audit log, and there are several built-in reports to analyze that log, as well as mechanisms to generate custom Excel reports.

**Workflow**

With Office SharePoint Server 2007 customers can build workflows that map to their important business processes. These capabilities enable more manageable collaboration, enforcable and measurable business processes, and more intelligent records management.

Microsoft Office SharePoint Server includes a set out-of-the-box workflows around approval, gathering feedback, gathering signatures, and others designed to map closely to the most often-used business processes in most organizations. Using either Sharepoint Designer or Microsoft® Visual Studio® 2005 new custom workflows can be created that codify crucial processes within the business.

## 4.3. Retaining & Archiving Spreadsheets

Spreadsheet archival is just one component of a larger records management process which includes the collection, management, and disposal of corporate records (information deemed important for the history, knowledge, or legal defense of a company) in a consistent and uniform manner based on company policies. The 2007 Microsoft Office System can help companies ensure that vital corporate records including critical spreadsheets are properly retained for legal, compliance, and business purposes and then properly disposed of when no longer needed. This section details the new capabilities around records management provided with Office SharePoint Server 2007.

### 4.3.1. Office SharePoint Server 2007 Record Repository

The core of the records management implementation in Office SharePoint Server 2007 is a stable, scalable, and efficient repository. Office SharePoint Server 2007 includes a specialized site template, known as the Records Repository, designed for records management.

The following capabilities, new to Office SharePoint Server 2007, help customers fulfill the requirements of records management.

**Vault abilities**

The Records Repository has several features that ensure the integrity of the files stored within it. First, it ensures that records are never automatically modified by the system. This means that records that are uploaded to a records repository and then downloaded again later will always be identical, byte for byte. Second, it has default settings that prevent direct tampering of records, by versioning any changes made to document contents and by auditing specific types of changes. Third, it allows records managers to add and maintain metadata on items separately from the record's metadata, so that information such as "who manages this item" can be changed without modifying the underlying record.

**Information management policies**

These policies provide controls that consistently and uniformly enforce the labeling, auditing, and expiration of records. Policies can be configured for a specific storage location or content type.

**Hold**

The Records Repository lets IT, records managers and legal authorities apply one or more holds that suspend the records management policies on items to ensure that they are kept unchanged during litigation, audits, or other investigations.

**Record Collection Interface**

Records repositories provide a set of services that aid in content collection. They let people and automated systems easily submit content to a records repository without necessarily having access or permission to any of the contents of the site.

**Record routing**

When content is submitted to a records repository, it can be routed to its proper location within the records management system based on its content type.

## 4.4. Developing Robust Spreadsheet Models

The capabilities of Microsoft Office Excel can be used to create a robust spreadsheet model that meets compliance challenges, and enhances productivity. The following capabilities in Microsoft Office Excel 2007 can help an organization deploy spreadsheet models that make it easier to become, and stay, compliant.

1. Cell Styles
2. Lock important cells
3. Tables as a native structure
4. Defined Names
5. Formula auditing tools

### 4.4.1. Cell Styles

Complex spreadsheets with multiple contributors can lack clarity and readability. Users interpret the spreadsheet's information differently, make errors based on assumptions, and are unable to quickly interpret or analyze the data. Cell formatting is an important tool that can be used to visually clarify the structure of a spreadsheet with color, font, borders, and data formats. Excel 2007 allows users to quickly define reusable cell formatting styles that make it easy to clearly indicate input cells, formulas, outputs and other key components. Changes made to the style are automatically applied to all cells using that style to making formatting updates simple. The resulting spreadsheet easier to read and less error prone.

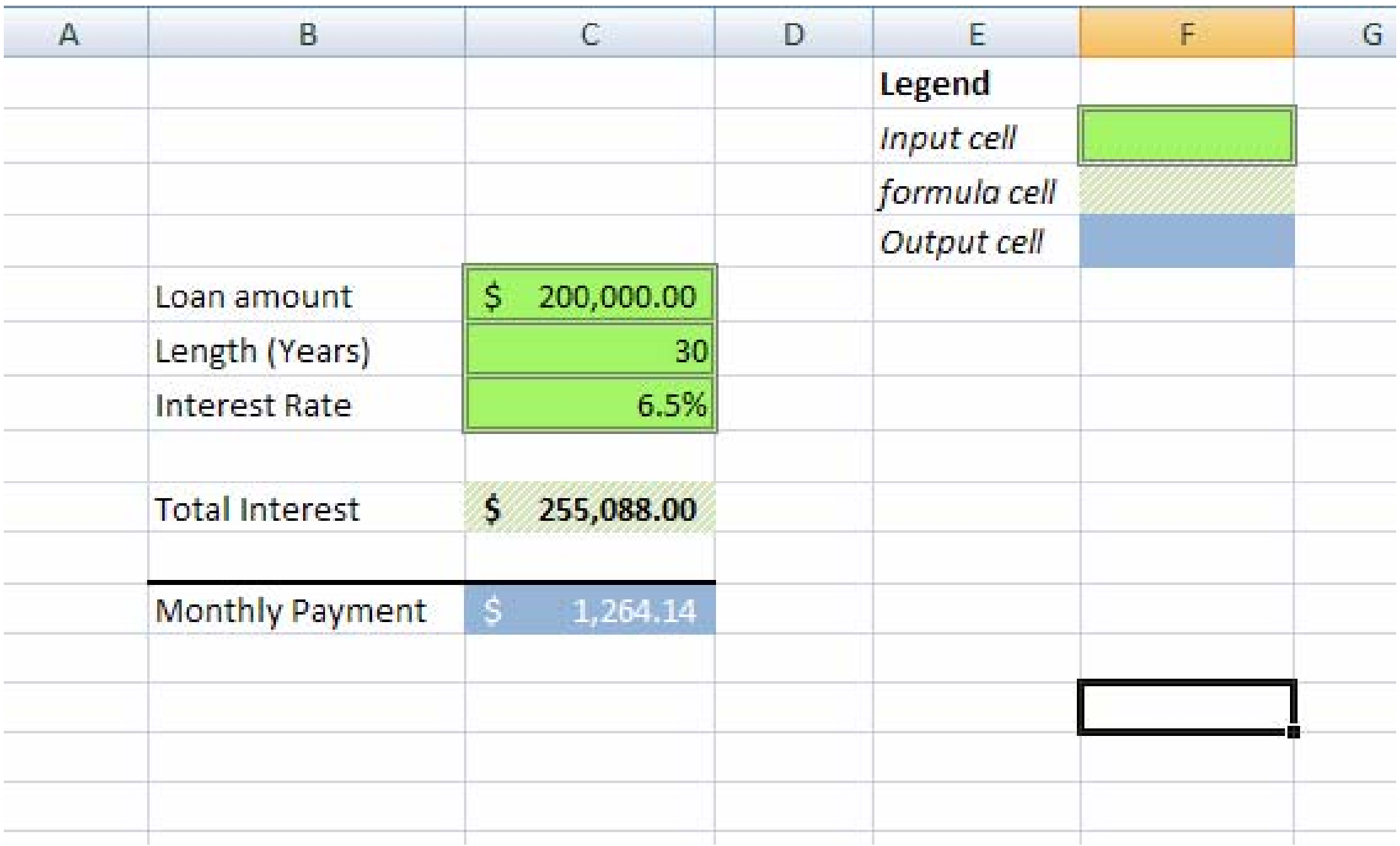

*Cell styles help distinguish inputs from calculation cells*

### 4.4.2. Lock important cells

In addition to making the spreadsheet more understandable. Organizations can reduce user errors by password protecting (or locking) specific cells, ranges, or sheets. This is a key step in the development of a robust spreadsheet.

**Protect Sheet**

Protect sheet allows an author to password protect selected cells within the worksheet, as well as prevent different types of changes to cells and other elements in the worksheet. This feature can be used to lock important areas of a spreadsheet preventing users from modifying the values or formulas in those cells.

**Allow users to Edit Ranges**

Similar to Protect sheet this command allows customers to lock down specific areas of a spreadsheet. With this feature however, an author can grant edit permissions to specific groups, users, or computers based on Windows NT authentication.

### 4.4.3. Tables as a native structure

One of the most common elements found spreadsheets today is the table, it is the standard method for organizing and displaying structured data. Excel 2007 now recognizes Tables as a native object in Excel spreadsheets, enabling users to create robust tables that maintain their structure and are significantly easier to interact with.

| | ProductName | UnitPrice | ExtendedPrice | Freight |
|---|---|---|---|---|
| Header Row — | ProductName | UnitPrice | ExtendedPrice | Freight |
| | Queso Cabrales | $ 14.00 | $ 168.00 | $ 32.38 |
| | Singaporean Hokkien Fried Mee | $ 9.80 | $ 98.00 | $ 32.38 |
| Data Region — | Mozzarella di Giovanni | $ 34.80 | $ 174.00 | $ 32.38 |
| | Manjimup Dried Apples | $ 42.40 | $ 1,696.00 | $ 11.61 |
| | Jack's New England Clam Chowder | $ 7.70 | $ 77.00 | $ 65.83 |
| | Manjimup Dried Apples | $ 42.40 | $ 1,261.40 | $ 65.83 |
| | Louisiana Fiery Hot Pepper Sauce | $ 16.80 | $ 214.20 | $ 65.83 |
| Total Row — | **Total** | **$ 167.90** | **$ 3,688.60** | **306.24** |

*A Table consists of three pieces:*

The Table feature increases productivity for many common tasks in Excel, such as:

- Manual data entry
- Copy / paste or drag / drop of data
- Sorting / Filtering data
- Selecting, navigating and scrolling around data

Tables also make more advanced tasks both easier to perform and more robust. As data is added to a Table, anything that is associated with that Table automatically adjusts in sync with the Table. Formatting will apply to new rows and Formulas will update to include new data. Charts, Pivot Tables, Conditional Formatting, and Data Validation will also all update to fit the new data.

Formulas that reference the data in a Table can do so by name (the name of the column, e.g. "Sales") rather than by an undecipherable A1-style address (e.g. D1:D10). This type of referencing is called "Structured Referencing" and increases the readability of formulas making them easier to maintain and edit later.

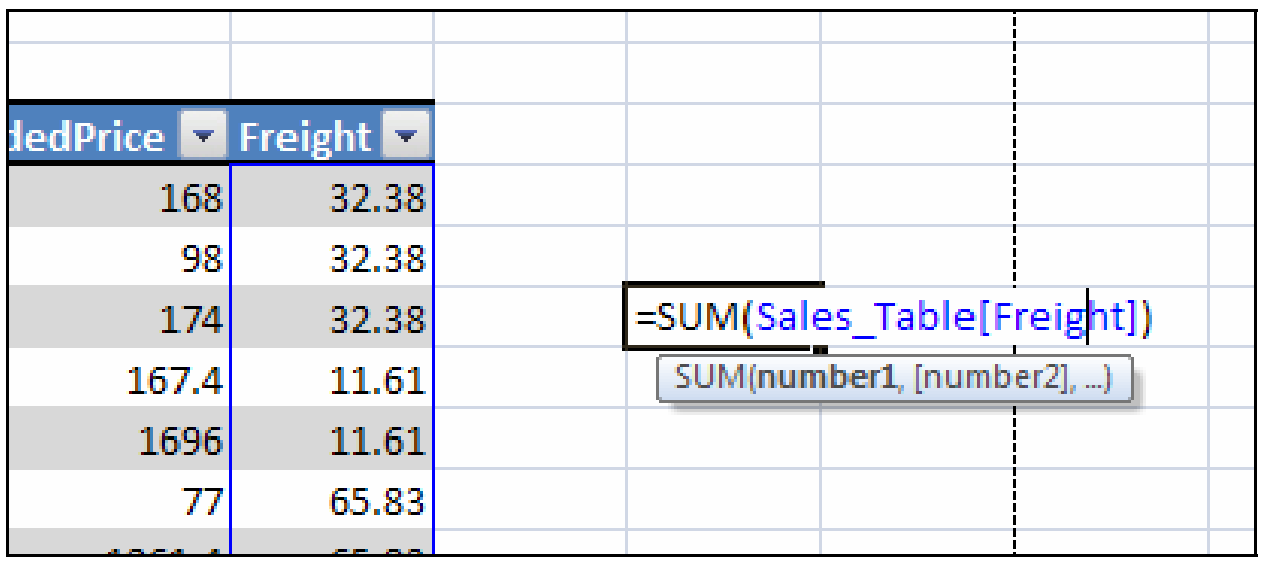

*Structured referencing in Excel 2007*

With Excel 2007 Tables, formatting features behaves in intelligent ways. For example, if alternate-row formatting is enabled on a Table, Excel 2007 will maintain the alternating format rule through actions that would have traditionally disrupted this layout, such as filtering, hiding rows, or manual rearranging of rows and columns. Additionally Excel 2007 ships with a large number of professionally designed table styles that look beautiful out of the box.

| | | | | |
|---|---|---|---|---|
| 17 | 1 | 0 | 17 | 8.53 |
| 15 | 2 | 0.01 | 29.7 | 8.53 |
| 7.75 | 4 | 0 | 31 | 8.53 |
| 13 | 2 | 0 | 26 | 8.53 |
| 13 | 2 | 0 | 26 | 9.54 |

*Excel 2007 tables support complex row and column banding that automatically adjusts with the data*

### 4.4.4. Defined Names

Defined names have always simplified writing formulas in complex spreadsheets, especially those shared among several people. However, when a spreadsheet contains hundreds or even thousands of defined names, it becomes more challenging to perform tasks like deleting multiple names, renaming names, and finding broken names.  The new Name Manager dialog is designed specifically for viewing and managing the names in a spreadsheet makes it much easier to do the following tasks:

- View important details including the name's reference, value, and scope.
- Create and scope Names.
- Rename existing names
- Delete multiple names at once
- Sort and Filter the Name List by common criteria including scope, type, and if the name returns an error.

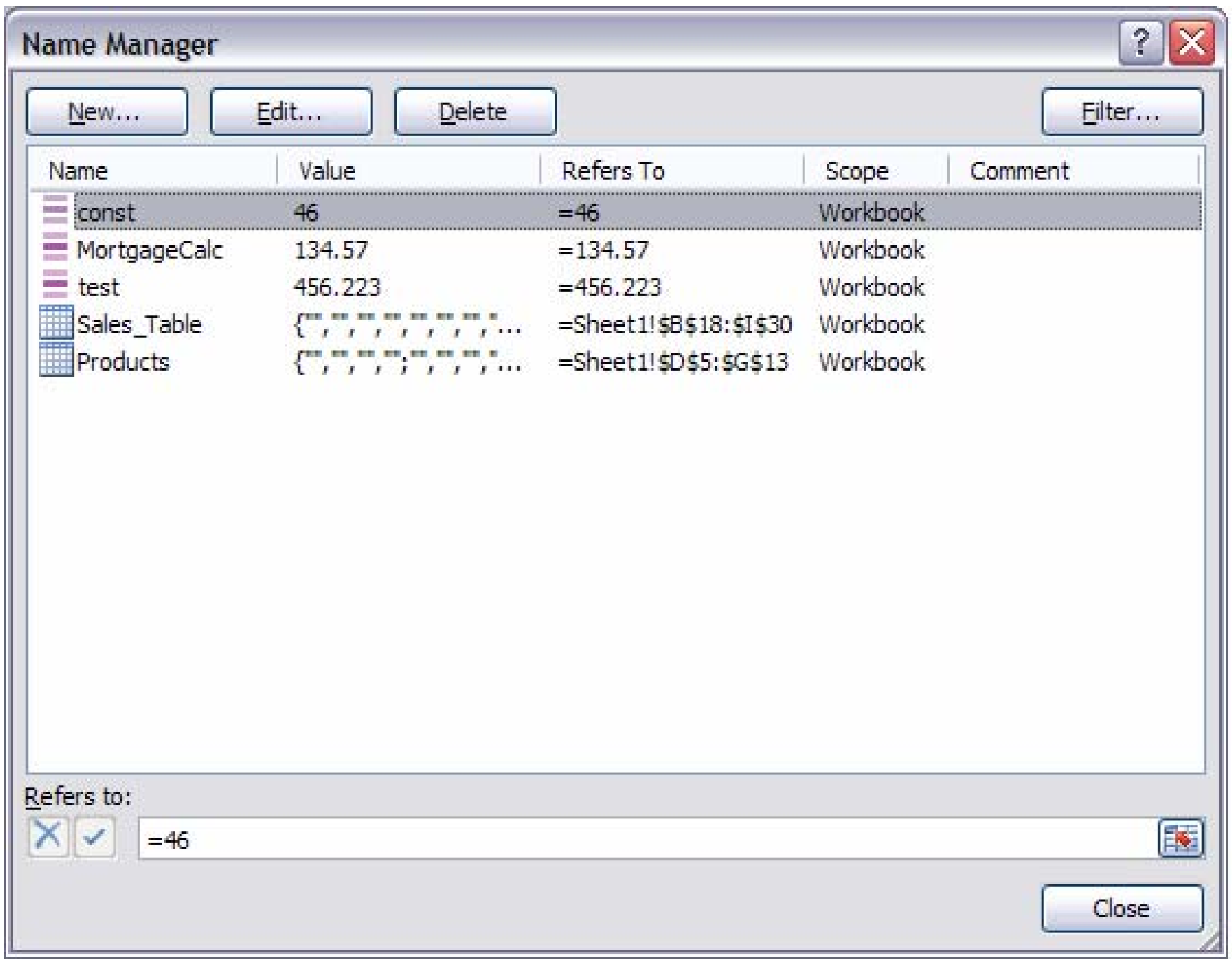

*The Name Manager dialog*

### 4.4.5. Formula Auditing tools

Regulatory compliance legistlation requires auditable and transparent practices for spreadsheets used in financial reporting. Excel 2007 provides auditing tools that, along with a consistent use of cell styles and naming conventions, can accelerate the testing of spreadsheet models, and reduce the risk of error once a spreadsheet is in production. Auditing tools in Excel 2007 provide the ability to:

- graphically display (or "trace") the relationships between cells and formulas
- trace a cell's precedents (the cells that provide the cell's data)
- trace a cell's dependents (the cells that depend on the cell's value)
- check for errors in a formula

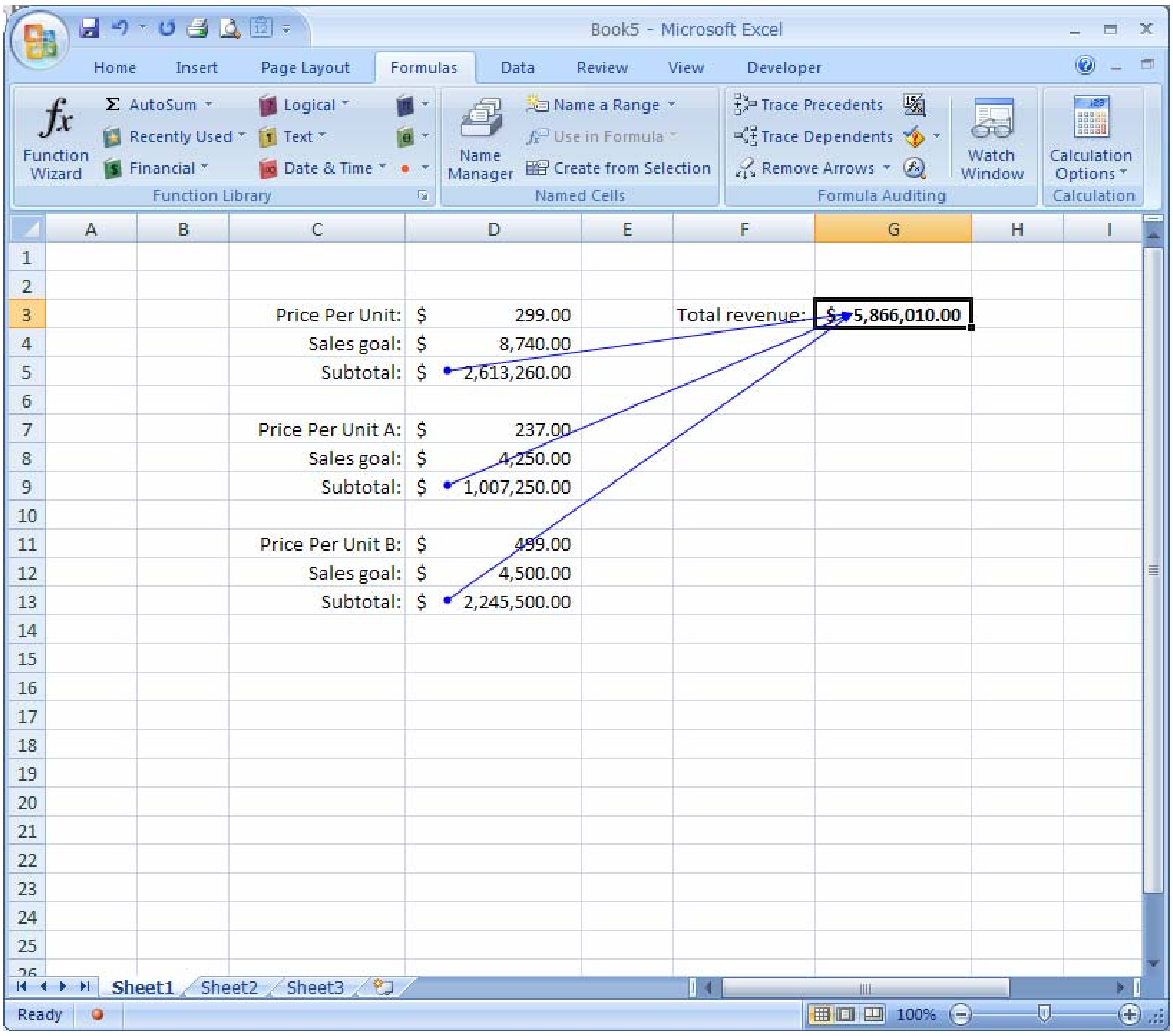

*A formula showing its precedents using auditing arrows*

## 5. CONCLUSION

Spreadsheets are commonly used as a critical resource in most organizations, yet they often receive little budgetary resources or sound management policies. This can result in an unnecessary exposure to regulatory compliance risks. As a result it is important for organizations to develop a spreadsheet compliance framework that includes rigorous process controls around the development, testing and use of business-critical spreadsheets. When these controls are combined with the current and forthcoming capabilities in the 2007 Microsoft Office System, companies will have greater success in implementing and enforcing spreadsheet policies.

## 6. LINKS

### Information about the 2007 Microsoft Office System

2007 Microsoft Office system website
http://www.microsoft.com/office/preview/default.mspx

Excel 2007 Blog
http://blogs.msdn.com/excel/

---

[i] PricewaterhouseCoopers. (2004), "The Use of Spreadsheets: Considerations for Section 404 of the Sarbanes-Oxley Act"

[ii] Panko, Raymond R., Ordway Nicholas. (2005) "Sarbanes-Oxley: What about All the Spreadsheets?" University of Hawaii.

[iii] Adapted from the waterfall software development lifecycle, a description can be found at http://en.wikipedia.org/wiki/Waterfall_model. 11/13/05